# XPS on Li Battery Related Compounds: Analysis of Inorganic SEI Phases and a Methodology for Charge Correction

*Kevin N. Wood[a] and Glenn Teeter[a,\*]*

[a]National Renewable Energy Laboratory, Golden, Colorado 80401

ABSTRACT

Accurate identification of chemical phases associated with the electrode and solid-electrolyte interphase (SEI) is critical for understanding and controlling interfacial degradation mechanisms in lithium-containing battery systems. To study these critical battery materials and interfaces X-ray photoelectron spectroscopy (XPS) is a widely used technique that provides quantitative chemical insights. However, due to the fact that a majority of chemical phases relevant to battery interfaces are poor electronic conductors, phase identification that relies primarily on absolute XPS core level binding-energies (BEs) can be problematic. Charging during XPS measurements leads to BE shifts that can be difficult to correct. These difficulties are often exacerbated by the coexistence of multiple Li-containing phases in the SEI with overlapping XPS core levels. To facilitate accurate phase identification of battery-relevant phases (and electronically insulating phases in general), we propose a straightforward approach for removing charging effects from XPS data sets. We apply this approach to XPS data sets acquired from six battery-relevant inorganic phases including lithium metal ($Li^0$), lithium oxide ($Li_2O$), lithium peroxide ($Li_2O_2$), lithium hydroxide (LiOH), lithium carbonate (Li$_2$CO$_3$) and lithium nitride (Li$_3$N). Specifically, we demonstrate that BE separations between core levels present in a particular phase (e.g. BE separation between the O 1s and Li 1s core levels in Li$_2$O) provides an additional constraint that can significantly improve reliability of phase identification. For phases like Li$_2$O$_2$ and LiOH where the Li-to-O ratios and BE separations are nearly identical, x-ray excited valence-band spectra can provide additional clues that facilitate accurate phase identification. We show that *in-situ* growth of Li$_2$O on Li$^0$ provides a means for determining absolute core level positions, where are all charging effects are removed. Finally, as an exemplary case we apply the charge-correction methodology to XPS data acquired from a symmetric cell based on a Li$_3$S-P$_2$S$_5$ solid electrolyte. This analysis demonstrates that accurately accounting for XPS BE shifts as a function of current-bias conditions can provide a direct probe of ionic conductivities associated with battery materials.

INTRODUCTION

Safe, reliable and scalable approaches to energy storage are crucial for enabling widespread adoption of renewable energy technologies. Unfortunately, many Li-based materials used in both conventional and next generation Li-ion battery (LIB) devices are highly reactive, which creates many challenges. For example, interfacial decomposition reactions occur spontaneously when an electrode (e.g., Li metal) is brought into contact with an electrolyte, creating an initial solid-electrolyte interphase (SEI) layer.[1] The SEI can continue to evolve during cycling as uncontrolled side reactions occur, and in some cases SEI phases themselves might be redox-active.[2-4] The SEI can dramatically affect both battery performance and cycling stability.[5] Therefore, understanding the processes that lead to SEI formation and evolution is a necessary step toward engineering stable, long lasting, and safe LIBs.[6]

To address these challenges, much effort has focused on understanding the morphological and chemical evolution of the SEI. One common method for probing the chemical composition of the SEI in liquid-electrolyte based systems is gas chromatography.[7-9] In these studies, GC measurements detect gases that evolve during the formation and cycling of a battery, providing clues about interfacial reactions. Complementary approaches such as optical microscopy, scanning electron microscopy and X-ray diffraction provide direct information about the morphology and crystallinity of SEI phases present in cycled electrode materials, but provide no direct information on chemical bonding environments.[10-14] In an attempt to measure chemical changes occurring at electrode/electrolyte interface, other techniques like vibrational spectroscopy, neutron diffraction, and nuclear magnetic resonance (NMR) have recently been employed with some success.

XPS analysis is often used to corroborate observations of techniques including Raman and NMR, and has been frequently used in recent decades to identify SEI phases. A challenge is that correct identification of chemical bonding environments using XPS relies on accurately determining small shifts in core-level binding energies (BEs). A related issue is the complexity introduced into XPS chemical-state analysis by the coexistence of multiple similar or related phases. For example, the Li 1s core levels associated with numerous distinct Li-containing chemical compounds (including several listed in Table I) common to battery materials lie within 3 eV of each other. Moreover, many SEI phases (including $Li_2O$, $Li_2O_2$, LiOH and $Li_2CO_3$) contain both Li and O, and in some of these phases the Li-to-O ratio is identical. Further complicating XPS analyses is the fact that sample charging effects on insulating materials can shift core levels by several eV, and BE calibration procedures can vary from lab to lab.

Issues related to XPS BE calibrations for battery-related materials and other alkali metal-based phases have been the focus of several recent studies, revealing that reliable identification of Li-battery SEI phases with XPS analysis is often no trivial task.[15–19] An early *in situ* study of Na intercalation into $TiS_2$ demonstrated that XPS BE shifts can be used to separate the relative ionic and electronic contributions to cell voltage.[15] A more recent photoelectron spectroscopy (PES) study of SEIs on graphite and $Ni_{0.5}TiOPO_4$ cathodes by Maibach et al., demonstrated peaks shifts for phases associated with the SEI were correlated with the degree of lithiation, and hypothesized the existence of an electric dipole at the SEI/cathode interface[16]. A follow-on study probed these interfacial dipole effects in more detail, and also demonstrated BE shifts for a variety of cathode active materials that were correlated with the electrode open-circuit voltage, and with degree of lithiation.[17] Additionally, Oswald documented systematic XPS peak shifts in several materials, and correlated these shifts with the presence of metallic Li and Na, hypothesizing that electrostatic interactions with alkali metals might account for these variations[18]. A subsequent study used XPS depth profiling to probe lithiation in graphite anodes also revealed BE shifts that depended on the degree of lithiation, and recommended that implanted Ar from sputter-cleaning or -profiling can be used as a reference for calibrating the BE scale.[19]

Considering the difficulty and importance of accurate BE calibration for battery materials a literature survey was performed. Table I summarizes some reported XPS BE's for the phases discussed in this manuscript.

In this work, we present evidence that the wide variations in assigned XPS BEs (specifically with regard to the battery-materials literature) occur largely as a result of surface charging inherent to electronically insulating materials during photoemission measurements. This occurs because photoelectrons ejected from an electronically insulating sample leave behind a net positive charge, leading to a rigid shift of the entire XPS spectrum to higher BEs. Differential charging effects, which cause different regions or phases on insulating samples to charge to varying degrees, further complicate these issues. A common practice for dealing with sample charging during XPS measurements is to use an electron flood gun to neutralize surface charge. This approach is effective to a degree, but generally there is no unambiguous way to precisely compensate for surface charge with this method, and consequently XPS core levels on insulating samples might be shifted be several eV (positive or negative) relative to their true positions. This has led to the common practice of calibrating XPS BEs by shifting the lowest-BE carbon peak to 284.8 eV (or a similar value, typically 284.6–285 eV), representing C-C bonding associated with adventitious carbon contamination. This approach can create inaccuracies in the reported BE values of the relevant core levels if: i) the amount of C-C bonding is minimal or hard to detect (a likely problem for the reactive nature of battery materials); ii) another lower BE functionality is present in the C 1s spectrum (e.g., a carbide); or iii) differential charging effects exists across the surface (as might be observed in cycled battery samples). Also, as shown in a recent study, a buried interphase potential difference that increases with cycling has been observed that can cause quite large shifts—on the order of nearly 2 eV—in the C 1s peak associated with the SEI.[17] Furthermore, hydrocarbon species in the electrode at different stages of cycling (for example, those in a pristine electrode vs. those in an SEI) might not be the same, and therefore could have differing BEs. Therefore, for all of these reasons, BE calibrations that rely on shifting the C 1s peak to a fixed value, especially for battery materials, can lead to substantial sample-to-sample and lab-to-lab variability.

**Table I:** Reported XPS binding energies in literature

| Li Metal | | Li₂CO₃ | | | |
|---|---|---|---|---|---|
| Li 1s BE (eV) | Reference | O 1s BE (eV) | C 1s BE (eV) | Li 1s BE (eV) | Reference |
| 48.7 | 20 | 531.5 | 289.8 | 55.2 | 21 |
| 52.1 | 22 | 531.5 | 289.5 | 55.1 | 23 |
| 52.3 | 24 | 531.8 | 290 | 55.3 | 25 |
| 52.3 | 26 | 531.8 | 290.1 | 55.5 | 27 |
| 52.3 | 28 | 531.8 | 290.1 | 55.4 | 29 |
| 53.0 | 30 | 532 | 290 | 55.4 | 13 |
| 53.9 | 31 | 532 | 289.8 | 55.5 | 32 |
| 54.2 | 33 | | | | |
| 55.5 | 34 | | | | |

| Li₂O | | | LiOH | | |
|---|---|---|---|---|---|
| O 1s BE (eV) | Reference | Reference | O 1s BE (eV) | Li 1s BE (eV) | Reference |
| 528.3 | 54.0 | 24 | 531.1 | 54.5 | 13 |
| 528.4 | 54.1 | 35 | 531.3 | 54.9 | 21 |
| 528.5 | 53.7 | 26 | 531.5 | 55.1 | 36 |
| 528.6 | 53.6 | 13 | 531.7 | 55.3 | 37 |
| 528.8 | 53.8 | 32 | 531.8 | 54.9 | 38 |
| 529.3 | 54.9 | 31 | 532 | 55.3 | 24 |
| 530.5 | 55.5 | 33 | 532.1 | 55.3 | 26 |
| 530.6 | 55.6 | 36 | | | |
| 531.3 | 55.6 | 21 | | | |
| 531.9 | 57.0 | 34 | | | |

| Li₂O₂ | | | Li₃N | | |
|---|---|---|---|---|---|
| O 1s BE (eV) | Li 1s BE (eV) | Reference | N 1s BE (eV) | Li 1s BE (eV) | Reference |
| 530.9 | 54.4 | 39 | 395.2 | 55 | 40 |
| 531.1 | 54.5 | 13 | 395.5 | 54.8 | 41 |
| 531.2 | 55.6 | 32 | 395.6 | 54.7 | 42 |
| 531.5 | 54.6 | 43 | 396 | 55.1 | 44 |
| 531.7 | 55.0 | 45 | | | |
| 532.1 | 55.3 | 26 | | | |
| 533.0 | 56.0 | 46 | | | |
| 533.1 | 56.4 | 36 | | | |

Adding to the challenges of correctly identifying Li-battery relevant phases is the often-difficult task of preventing exposed Li-containing surfaces from reacting with ambient environments, including ultra-high vacuum (UHV) conditions. For example, when Li metal is transferred directly from a glovebox to an XPS chamber via an integrated load-lock system, metallic Li is generally not detectable. Only after extended periods of $Ar^+$ sputter-cleaning to remove adventitious surface contaminants can metallic Li be observed, and thereafter, reactions with background gases (primarily $H_2O$) will convert the surface to $Li_2O$[47–49]. Even under UHV conditions, it is only possible to keep highly reactive samples clean over time spans of minutes (at a base pressure of $1\times10^{-10}$ torr, one can expect ~0.1 monolayer of contamination to form in ~15 minutes), and rates of surface contamination even in the cleanest gloveboxes are significantly higher.

Despite these challenges, XPS is a very useful technique for battery analysis, due to its high surface sensitivity and the fact that it provides direct information on chemical bonding environments at battery interfaces. Nevertheless, the potential difficulties with accurate XPS chemical-state assignments can create confusion about which SEI phases are beneficial vs. detrimental to overall battery performance and stability. To address these issues, in this work we propose a method for analyzing XPS spectra acquired from samples where charging is an issue. Briefly, the proposed method outlined in this work for correcting XPS charging artifacts to facilitate accurate phase identification is as follows. First, it is necessary to determine the characteristic BE separations (ΔBEs) between the relevant core levels (e.g., the separation between the O 1s and Li 1s peaks for $Li_2O$) for specific phases that might be present. Subsequently, when performing curve fitting on spectra from a sample that is suspected to contain a particular set of phases, the relevant core levels should be constrained by the pre-determined ΔBE values, and at the same time the peak intensities should also be constrained according to the correct elemental ratios for each phase. (As stated previously, care must be taken with respect to the possibility that phases are not distributed homogeneously within the XPS detection volume. Differences in electron inelastic mean free path values can significantly alter effective elemental sensitivity factors. Unless accounted for properly, these effects could substantially distort curve-fitting results.) By constraining ΔBE values (rather than absolute BEs), this method is insensitive to BE shifts associated with charging phenomena often seen on electrically insulating materials, and is thereby less susceptible to phase assignment errors. In this study, this method is applied to specific inorganic Li-battery relevant phases including $Li^0$, $Li_2O$, $Li_2O_2$, $LiOH$, $Li_3N$, and $Li_2CO_3$, and ΔBE parameters are quantified for these materials. It is worth noting that this approach can be applied to any materials where charging during XPS measurements is an issue, not just battery materials.

EXPERIMENTAL

Li$_2$CO$_3$, Li$_2$O$_2$, LiOH and Li$_2$O were transferred through air and analyzed by a Kratos AXIS Nova XPS system. Base pressures were better than 2×10$^{-9}$ torr. 'Air-free' XPS measurements, for Li$_2$O, Li$_3$N, and Li metal-based samples, were performed using a glovebox-integrated Phi 5600 XPS system. Direct sample transfers of Ar-packaged materials to the XPS were completed in the attached glovebox under <10 ppm moisture and O$_2$ conditions. Base pressures for the 5600 system were below 7×10$^{-10}$ torr. Powder samples were pressed into indium metal foils, which were then mounted on XPS sample holders. Photoelectrons were generated using monochromatic Al Kα X-ray excitation (hν= 1486.7 eV). The spectrometer BE scale was calibrated by measuring valence-band and core-level spectra from sputter-cleaned Au, Ag, and Cu foils (E$_F$ = 0.00 eV, Au 4f$_{7/2}$ = 83.96 eV, Ag 3d$_{5/2}$ = 368.26 eV, and Cu 2p$_{3/2}$ = 932.62 eV).[50] Curve fitting and data processing were performed using Igor Pro with a custom program adapted from Schmid, et al.[51] The Li metal Li 1s core level exhibits an asymmetric Doniach-Sunjic line shape, characteristic of metallic conductors.[52] In this work a pseudo-Voigt function was employed to fit the asymmetric line shapes.[51] Valence-band maximum (VBM) values, which represent the characteristic binding energies of the most weakly bound occupied electronic states in a material, were extracted from valence-band spectra in the standard way by finding the intersections between least-squares best-fit lines representing the valence-band onsets and background counts. VBM and core-level BE uncertainties were derived from curve-fitting standard deviations (σ values): quoted core-level BE uncertainties represent ±3σ; and uncertainties for subsidiary ΔBEs values were calculated by propagating individual BE uncertainties. VBM uncertainties were calculated by propagating standard deviations associated with the individual straight-line fits associated with the valence-band onsets and backgrounds, respectively. Full-width at half-maximum (FWHM) values were constrained in curve fitting to < 2.0 eV; typical optimized values are summarized in the SI in Table SI1. Elemental ratios were calculated using tabulated sensitivity factors from the Handbook for X-ray Photoelectron Spectroscopy, published by Physical Electronics. While the specific materials analyzed here may have slightly different sensitivity factors than the tabulated values, the ratios appeared close to expected values. Dosing experiments were performed by leaking high purity oxygen (99.9%), nitrogen (99.99%) or H$_2$O vapor (from deionized water purified *in situ* by freeze-pump-thaw cycles) into the chamber typically at 1×10$^{-8}$ torr. Gas doses are measured in Langmuirs (L), where 1 L = 1×10$^{-6}$ torr·s.

RESULTS/DISSCUSION

To better understand trends in XPS peak assignments recorded in the battery literature, a review of reported BE assignments for eight different Li-containing phases was performed. This survey revealed that BE assignments for a particular core level in a given phase typically vary over ranges of several eV. As a point of reference, typical XPS BE uncertainties are < ±0.2 eV, so the observed range in BEs cannot be ascribed primarily to measurement uncertainties. To further elucidate these variations, for each compound the anion core-level BEs was plotted as function the Li 1s BEs, as illustrated in Fig. 1 for Li$_2$O. In this example, the O 1s core level in Li$_2$O has been assigned over a broad range, ~527–532 eV. Similarly, the Li 1s core level varies over the range ~53–57eV. Despite the striking range of BE assignments, a clear linear trend is observed. Least-squares fitting to a straight line (with slope = 1) provides a good fit to the data. The unity slope makes sense if the observed BE shifts result primarily from charging effects, in which case all core levels are expected to shift by the same amount. The y-intercept of the unity slope best-fit line provides the characteristic BE difference (ΔBE) between the Li 1s and O 1s core levels for $Li_2O$. This analysis also reveals several outlier data points that differ in some cases from the global fit by more than 0.3 eV, which is larger than one would expect based on typical XPS BE uncertainties. The existence of such outliers point to other potential sources of error, beyond just charging effects.

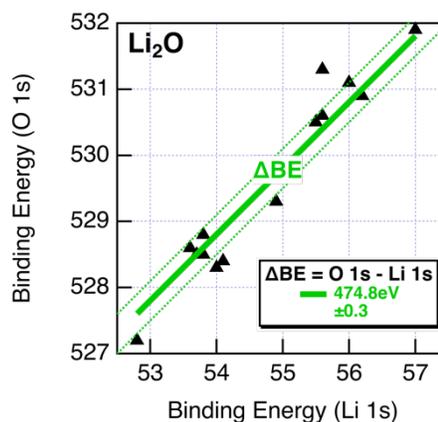

**Fig. 1:** *Plot comparing XPS BE assignments in literature for Li$_2$O. The solid green line shows the average binding energy separation (ΔBE) between peaks assigned to Li$_2$O in both the O 1s and Li 1s core levels.*

While the trend in Fig. 1 is clear, it is not obvious what the correct ΔBE value should be. For example, the lower four outlier data points are well fit by the unity slope line with ΔBE = 474.3 eV, substantially different than the global best fit ΔBE = 474.8 eV. Further complicating matters is the fact that many potential reference materials have phase impurities, especially within the information depth of XPS. This is the case for $Li_2O$ powder, which is known to convert to LiOH upon exposure to $H_2O$ vapor.[47–49] In our studies, XPS analysis of $Li_2O$ powder (99% purity, Sigma-Aldrich) revealed nearly complete conversion of $Li_2O$ to LiOH (spectra and analysis to be discussed in more detail later). Therefore, reliance on as-received materials, especially reactive or moisture-sensitive materials used in batteries, can lead to inaccurate phase assignments in XPS reference measurements.

Clearly what is needed is an absolute BE calibration for at least one Li-containing phase, which can be used to calibrate other phases. In the literature, Li metal ($Li^0$) has been assigned BE values over a 7-eV range. This is surprising, because the measured BE of a metallic sample should be unaffected by charging, and therefore one might expect that Li metal would show the least dispersion in assigned BEs. Possibly some of the variability can be traced to the extreme reactivity of Li metal. Even when stored in a glovebox, metallic Li reacts quickly with trace amounts of residual gases. This leads to surface contamination layers comprised of phases including $Li_2CO_3$, $Li_2O$ and LiOH (supporting information; Fig. SI1). Therefore, to achieve the cleanest possible $Li^0$ surface, in this study Li foil was sputter cleaned with 3-keV $Ar^+$ ions for ~3 h. Subsequently, XPS spectra were acquired, while sputter-cleaning continuously, to obtain a low-noise baseline XPS spectra for $Li^0$, with minimal surface contamination, as shown in Fig. 2. Similar XPS spectra acquired without continuous sputtering revealed significantly higher levels of surface contamination (supporting information; Fig. SI2). As evident from the baseline $Li^0$ spectra, even with continuous sputter cleaning, a slight amount of $Li_2O$ still exists on the surface (phase composition: 1.7%; O concentration: 0.6%). However, this slight degree of contamination does not interfere with analysis of the Li⁰ core level. The Li 1s core level in Fig. 2 is characterized an asymmetric line shape peaked at 54.97±0.06 eV, and there is a broad plasmon-loss feature between 66–60 eV (Shirley and coworkers have also demonstrated another plasmon-loss present around 70 eV [53]). In general, this Li 1s peak position is high compared to the average literature assignment (53.0 eV). However, we can decisively rule out the possibility that surface contamination or charging affected this BE assignment is based on two factors: i.) the absence of significant levels of surface contamination that might affect measured chemical states; and ii.) the clear existence of spectral features associated with metallic Li, including the plasmon-loss feature (Fig. 2) and the metallic Fermi edge in the valence-band spectrum (Fig. 2).

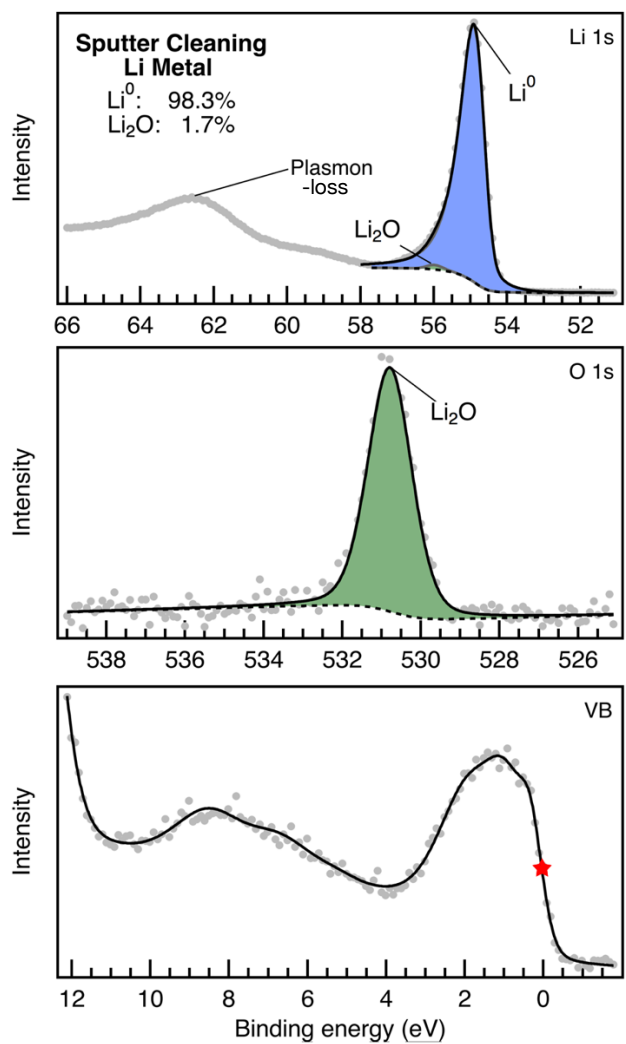

**Fig. 2:** *High-resolution XPS spectra of sputter cleaned metallic lithium. Even under continuous sputtering a small amount of oxide remained on the surface. This oxide has spectral features associated with $Li_2O$ (discussed later). Therefore, it must be fit in both the Li 1s (difficult to see) and the O 1s spectra. In the Li 1s panel, the $Li_2O$ peak is shaded green and the $Li^0$ is shaded blue.*

It should be noted that while we are confident the Li⁰ BE position is accurate and non-ambiguous due to the above-mentioned effects, a charging artifact was still observed on metallic Li foil samples. For the specific instrument used to acquire the data in Fig. 2, a standard 4-point BE calibration measurement (using the Au $4f_{7/2}$, Cu $2p_{3/2}$, and Ag $3d_{5/2}$ core levels, and the metallic Fermi edges, acquired from sputter-cleaned foils) was applied, such that $E_F = 0$ eV on the BE scale. However, even after this calibration had been applied, it was observed that the Fermi edge on sputter-cleaned Li metal was offset from zero by +0.12 eV. This slight discrepancy is consistent with a slight buildup of positive charge on the sample. To confirm this effect was due to charging, the electron flood gun was used to supply excess electrons to the surface during XPS measurements. This shifted Li⁰ Fermi level to 0.01eV, well within error of the values measured for the calibration samples (Fig. SI4). Therefore, even with sputter cleaned Li⁰ charging artifacts were observed. We attribute this effect to non-negligible contact resistance between the Li foil and the sample holder. Specifically, even though the back of the Li foil was mechanically abraded prior to tightly clamping it to the XPS sample holder, we hypothesize that a resistive $Li_2O$ layer forms between the Li metal and the metallic sample holder. Possibly this Li$_2$O layer originates with the Li foil or it might form from reactions between metallic Li and surface oxides (mainly CrO$_2$) on the 316 SS sample holder.

To circumvent the challenges of correctly identifying and assigning BE positions for bulk powder reference samples (e.g. Li$_2$O, LiOH, etc.) that can have large amounts of surface contamination and/or serious charging effects, we prepared sputter-cleaned Li metal surfaces and performed *in situ* gas dosing with O$_2$, H$_2$O, and N$_2$. In the first set of measurements, after sputter-cleaning for 3 h the sample was dosed with O$_2$ at $1\times10^{-8}$ torr. XPS measurements were performed during gas dosing (Fig. 3) to document changes to surface composition, chemical states and the valence band. Prior to dosing, a pristine Li$^0$ peak is observed in the Li 1s. After a dose of 3 L a new chemical state appears at higher BE; concurrently a peak grows in the O 1s spectrum centered near 531 eV. Upon continued dosing an isosbestic point emerges in the Li 1s spectra, consistent with the co-existence of two phases which change in relative concentration as the net O$_2$ dose increases. These observations, combined with elemental ratios, demonstrates unsurprisingly that O$_2$ exposure converts Li$^0$ to Li$_2$O. These O$_2$-dose dependent spectra therefore provide an ideal reference for Li 1s and O 1s BEs associated with Li$_2$O. As the Li$_2$O layer grows a slight shift to higher BE is observed for both the O 1s peak and the Li 1s feature associated with Li$_2$O. Likely this results from subtle evolution of chemical states associated with the Li$^0$/Li$_2$O interface, which remains within the XPS information depth, with a more pronounced effect on line shapes at early times. On the other hand, for O$_2$ doses > 30 L, the Li 1s feature associated with Li$^0$ is completely attenuated, indicating that the final spectrum in the series represents pure Li$_2$O. On this basis, we determine that the BE positions for pure Li$_2$O are 56.40±0.06eV and 531.20±0.06eV for the Li 1s and O 1s core levels, respectively. Furthermore, analysis of Fig. 3(c) plainly shows dramatic changes in the valence-band spectrum as the surface converts to Li$_2$O. Specifically, the metallic Fermi edge disappears, the valence-band minimum (VBM) shifts to 4.69±0.09eV, and a broad peak grows in between 5–9 eV.

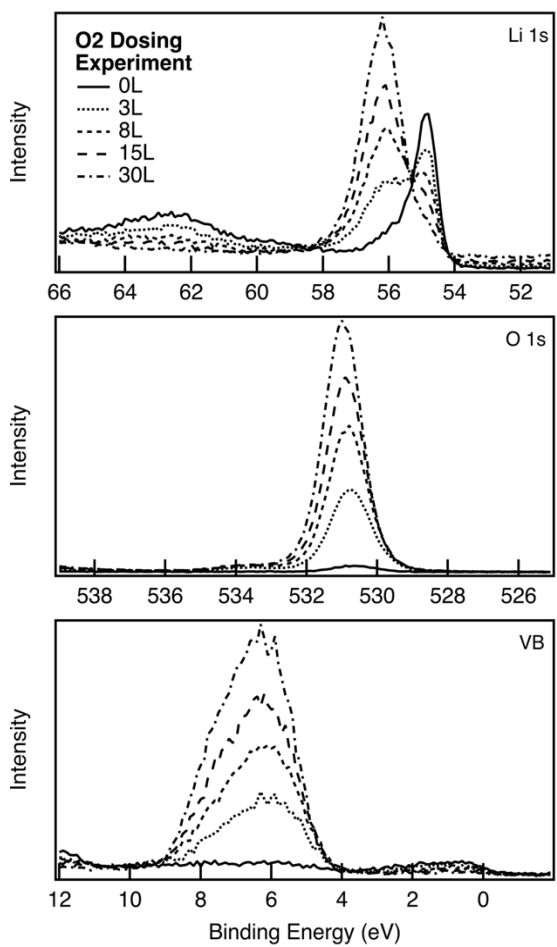

**Fig. 3:** *High-resolution XPS spectra for (a) Li 1s, (b) O 1s core levels, and (c) the valence band during in-situ O$_2$ dosing on sputter-cleaned Li metal, documenting formation of Li$_2$O.*

It is interesting to note that as the Li$_2$O feature grows in the Li 1s spectra in Fig. 3, the overall peak intensity appears to actually decrease, somewhat counterintuitively, since the relative abundance of Li atoms in Li$_2$O is only 2/3 that in metallic Li. A likely explanation for this result is related to the strong plasmon loss features that can be observed in metallic Li and other alkali metals, which are not observed in Li$_2$O. In Fig. 3, one can see that there is a significant intensity in the first plasmon-loss feature centered at ~62.5 eV; and additional plasmon loss features can be observed at higher BEs (not shown in Fig. 3). Therefore, since a high proportion of Li 1s photoelectron intensity is lost to the creation of plasmons, apparently the effective elemental sensitivity factor for Li in metallic lithium is significantly lower than in Li$_2$O and other Li-containing phases.

As mentioned previously, in typical battery samples numerous Li-based phases are likely to coexist, and many samples (especially those that have come in contact with an electrolyte) are likely to charge when analyzed. Therefore, characteristic BE separations like the one demonstrated in Fig. 1 can facilitate accurate phase identification. For Li$_2$O (Fig. 3), the O 1s – Li 1s separation is ΔBE = 474.80±0.09eV. In general, valence-band features can be helpful to further refine phase assignments, especially in cases where a combination of elemental ratios and core-level BEs, or BE separations, produces ambiguous results. In the case of Li$_2$O, the O 1s – VBM value is ΔBE = 526.51±0.11eV.

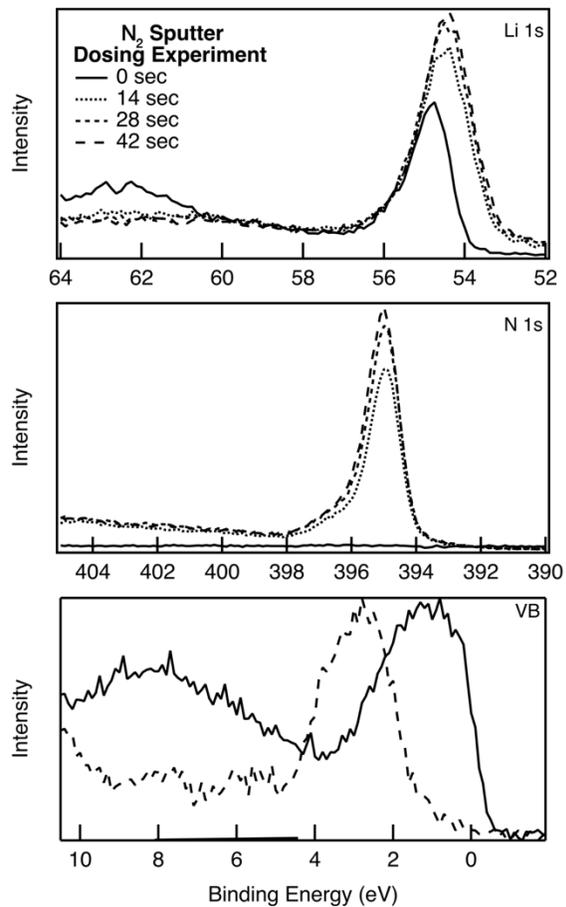

Similar dosing experiments on clean metallic Li were performed using H$_2$O vapor and N$_2$. The goal of the H$_2$O dosing experiments was to produce LiOH surface layers, but under the accessible experimental conditions (p$_{H2O}$ < 1×10$^{-6}$ torr and net dose < 1×10$^4$ L) only Li$_2$O formation was observed. In the case of N$_2$, even after N$_2$ doses >50 L (at 1×10$^{-8}$ torr) no changes were observed (Supporting information; Fig. SI5). However, the use of a sputter-ion source to impinge dissociated, ionized N$^+$ species (1 keV incident energy) on the Li metal surface resulted in rapid incorporation of N. As shown in Fig. 4, after as little as ~40 s of exposure to energetic N$^+$ ions, the Li$^0$ peak is almost

**Fig. 4:** *XPS spectral data showing the formation of Li$_3$N via energetic ion bombardment.*

completely gone, evident by the disappearance of both the Li$^0$ plasmon peak and the metallic Fermi edge. It is important to note that for the 1-keV incident kinetic energy used, the N$^+$-ion penetration depth is expected to be quite limited, on the order of several nm at most. Peak fitting of the Li 1s and N 1s spectra reveals an elemental ratio at the surface close to the expected $Li_3N$ value of 3:1 (77% Li to 23% N), confirming the presence of $Li_3N$. XPS depth profiling (supporting information,

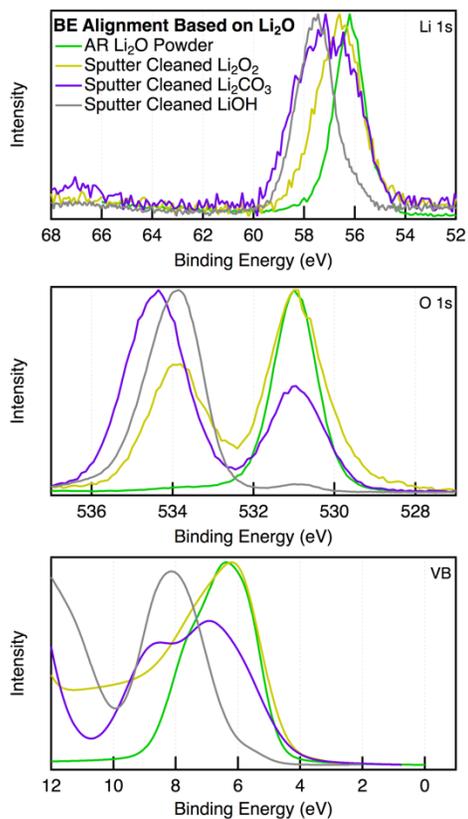

Fig. SI6) showed that this ratio was observed over a depth of several hundred nanometers, well beyond the information depth of XPS. The substantial overall thickness of $Li_3N$ indicates that once nitrogen is incorporated into the Li-foil substrate, interdiffusion is sufficiently facile to enable continued growth of the $Li_3N$ layer. We determine $Li_3N$ BE values to be 54.56±0.06eV, 395.33±0.06eV, and 1.38±0.15eV for the Li 1s and N 1s core levels, and the VBM, respectively. The N 1s – Li 1s separation is therefore $\Delta BE = 340.77\pm0.09$eV, and the N 1s – VBM separation is $\Delta BE = 393.95\pm0.16$eV.

**Fig. 5:** *XPS spectra of various Li compounds where $Li_2O$ was intentionally formed on the surface via $Ar^+$ sputtering to enable an accurate BE calibration. The valence-band spectra have been smoothed to improve clarity; the unsmoothed spectra are also shown in Fig. 6.*

Unfortunately, not all Li-containing SEI phases can be grown by vapor-phase dosing of metallic Li. Therefore, accurate assignment of BE values for XPS spectra acquired from readily available reference materials are also needed. However, surface reactions of these materials with adventitious C-containing species, $O_2$ and $H_2O$ make accurate identification challenging. Analysis of commercially available Li$_2$O powder demonstrated that by the time of analysis nearly all Li$_2$O in the near-surface region had been converted into LiOH and Li$_2$CO$_3$ (as shown in Fig. 6) with only ~5% Li$_2$O remaining. These reactions are expected and has been documented elsewhere.[49] Fig. 6 illustrates that Li$_2$CO$_3$ in particular tends to be a major component of any sample that has experienced a significant exposure to the ambient atmosphere. The Li$_2$O powder sample was also insulating, and required charge compensation via electron flood gun (Fig. SI4). Further complicating matters is the fact that there is no Fermi edge in valence-band spectra from non-metallic materials, which limits the utility of valence-band spectra for direct BE calibrations in these types of samples. On the other hand, the availability of calibrated Li 1s and O 1s BE values from $Li_2O$ grown in situ on Li°, as described above, can facilitate calibrations on other Li-containing phases. In cases where Li₂O can be identified by XPS as a secondary phase at the surface of a commercially available reference material, the Li 1s and O 1s associated with the Li₂O phase can provide a rigorous internal BE calibration for that sample.

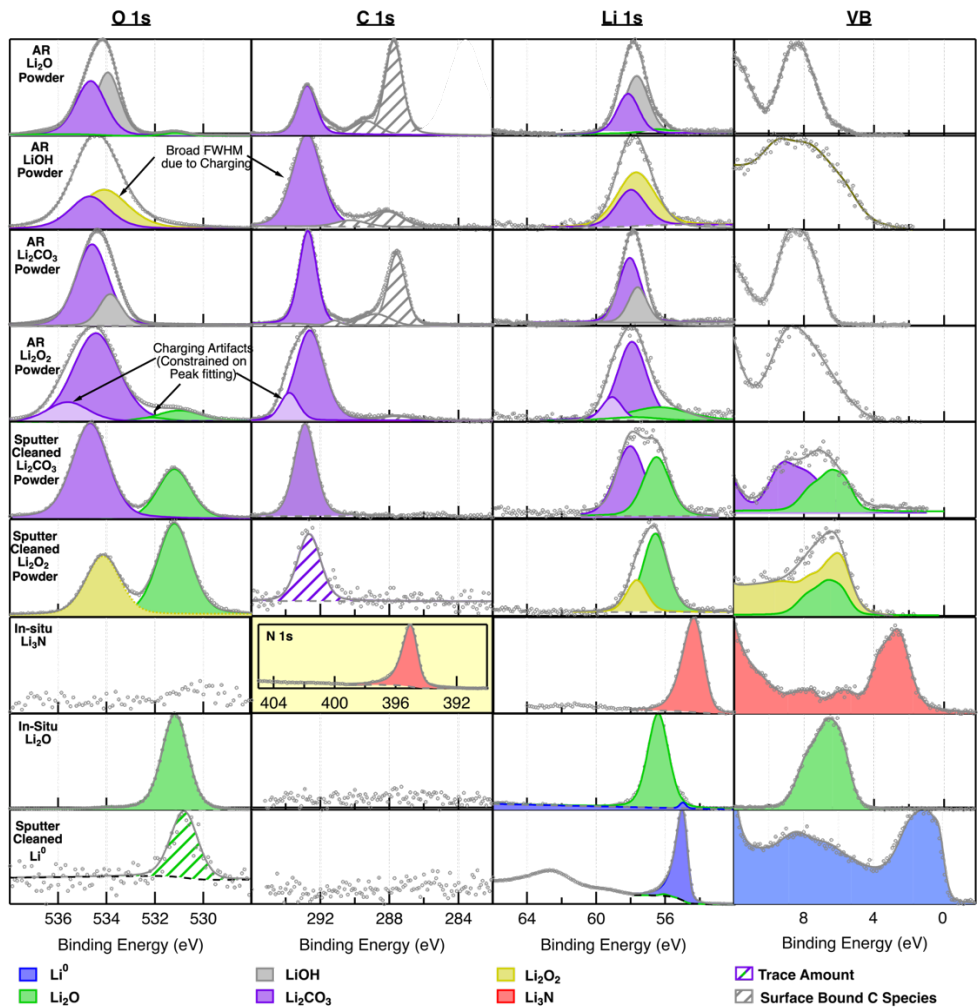

**Fig. 6:** *Comparison of XPS data sets from four phases containing both Li and O, acquired from as-received (AR) commercially available powders ($Li_2O$, LiOH, $Li_2CO_3$, $Li_2O_2$). Also shown are spectra from $Ar^+$-ion sputter cleaned $Li_2CO_3$, $Li_2O_2$, and from $Li_2O$ formed via in-situ oxidation of metallic lithium. Spectra from sputter-cleaned $Li^0$ are provided for reference. A small $Li^0$ feature is seen in the $Li_2O$ spectra from the underlying metallic Li surface. In general, air-exposed powder reference samples included significant amounts of secondary phases in the near-surface region, i.e., within the XPS detection volume. In particular, features associated with $Li_2CO_3$ are seen in spectra from as-received Li2O and LiOH; and LiOH is also present at the surface of the $Li_2O$ powder.*

Fortuitously, Ar⁺ ion sputter-cleaning leads to formation of the highly stable phase Li₂O at the surface of many Li-containing compounds. This phenomenon provides a convenient means for determining absolute BE positions for Li₂O₂, LiOH, and Li₂CO₃, and potentially other Li-containing phases. In the present study, the as-received powders (including Li₂O₂, LiOH, and Li₂CO₃) were first analyzed by XPS to determine the baseline spectra. Then the powder reference samples were sputter-cleaned for up to 15 s. Depending on the sample, a portion of the surface (5–70%) was converted to Li₂O. The BE scales for core-level spectra from each sample were then adjusted so that the Li 1s feature in Li₂O shifted to 531.2 eV. The Li₂O-shifted spectra are shown in Fig. 5. For each of these materials curve fitting was used to verify elemental ratios. The spectra in Fig. 5 show that peaks associated with Li₂O, and with the phases of interest (i.e., Li₂O₂ Li₂CO₃, and LiOH), are well separated in the O 1s spectra. After alignment the close similarities between Li₂O₂, Li₂CO₃, and LiOH in both the Li 1s and O 1s core levels become evident. In fact, based on these similarities LiOH and Li₂O₂ are virtually indistinguishable. For Li₂CO₃, both the O 1s and Li 1s are peaks are shifted to slightly higher BEs than the corresponding peaks in LiOH and Li₂O₂. In cases where BEs and ΔBE values are inconclusive, valence-band spectra can provide additional information to aid in phase assignments. As can be seen in Fig. 6, the valence-band features for each phase are significantly different. Furthermore, using the BE separation between a relevant core level and the VBM can be very helpful in situations where charging is an issue, as is generally the case for all phases studied in this work. A summary of the calibrated BE reference assignments, BE separations, and VBM values for all phases investigated in this study are provided in Table II. XPS spectra for the as-received $Li_2O$, $Li_2O_2$, LiOH, $Li_2CO_3$, and Li metal reference samples, with $Li_2O$-calibrated BE scales, are plotted in Fig. 6.

**Table II:** *Values determined in this study for both the absolute BE and ΔBE values for various inorganic Li-containing phases. $Li^0$ values were determined using sputter cleaned Li metal. $Li_2O$ and $Li_3N$ were samples were formed by in-situ oxidation and nitridation of Li metal foil, respectively. The LiOH parameters were extracted by correlating data sets measured on LiOH and $Li_2O$ powders (LiOH forms adventitiously on the surface of air-exposed $Li_2O$, as can be seen in Fig. 6). BE parameters for $Li_2O_2$ and $Li_2CO_3$ were determined from sputter-cleaned $Li_2O_2$ and $Li_2CO_3$ powders, respectively.*

| Absolute Binding Energy Positions | | | | | | | | | | | | | | |
|---|---|---|---|---|---|---|---|---|---|---|---|---|---|---|
| $Li^0$ | | | $Li_2O$ | | | LiOH | | | $Li_2O_2$ | | | $Li_2CO_3$ | | | $Li_3N$ | | |
| | BE (eV) | 3σ (eV) | | BE (eV) | 3σ (eV) | | BE (eV) | 3σ (eV) | | BE (eV) | 3σ (eV) | | BE (eV) | 3σ (eV) | | BE (eV) | 3σ (eV) |
| Li 1s | 54.97 | ±0.06 | Li 1s | 56.4 | ±0.06 | Li 1s | 57.40 | ±0.07 | Li 1s | 57.51 | ±0.08 | Li 1s | 58.05 | ±0.11 | Li 1s | 54.56 | ±0.06 |
| O 1s* | 530.97 | ±0.07 | O 1s | 531.2 | ±0.06 | O 1s | 533.77 | ±0.07 | O 1s | 534.15 | ±0.07 | O 1s | 534.67 | ±0.07 | N 1s | 395.33 | ±0.06 |
| | | | | | | | | | | | | C 1s | 292.89 | ±0.06 | | | |
| VBM | 0 | ±0.13 | VBM | 4.69 | ±0.09 | VBM | 6.01 | ±0.11 | VBM | 3.33 | ±0.37 | VBM | 5.75 | ±0.59 | VBM | 1.38 | ±0.15 |
| *At least small amount of oxygen is always present in form of Li rich lithium oixde | | | | | | *Very close O 1s - Li 1s means that O 1s-VB should be used to tell between $Li_2O_2$ and LiOH | | | | | | | | | | | |
| | ΔBE | 3σ (eV) | | ΔBE | 3σ (eV) | | ΔBE | 3σ (eV) | | ΔBE | 3σ (eV) | | ΔBE | 3σ (eV) | | ΔBE | 3σ (eV) |
| O 1s -to- Li 1s | 476.00 | ±0.10 | O 1s -to- Li 1s | 474.80 | ±0.09 | O 1s -to- Li 1s | 476.37 | ±0.10 | O 1s -to- Li 1 | 476.64 | ±0.11 | O 1s -to- Li 1s | 476.62 | ±0.12 | N 1s -to- Li 1s | 340.77 | ±0.09 |
| Li 1s -to- VBM | 54.97 | ±0.14 | O 1s -to- VBM | 526.51 | ±0.11 | O1s -to- VBM | 527.76 | ±0.13 | O 1s -to- VBM | 530.82 | ±0.38 | O 1s -to- C 1s | 241.78 | ±0.09 | N 1s -to- VBM | 393.95 | ±0.16 |
| | | | | | | | | | | | | C1s -to- Li 1s | 234.84 | ±0.12 | | | |

To demonstrate how the BE assignments presented in this work compare with values in the literature, Fig. 7 summarizes reported values for six Li-containing phases in the same manner as

Fig. 1. Fig. 7 then compares these literature values to absolute BE assignments (yellow/red circles) and ΔBE separation data (solid lines) reported in this study. In Fig. 7 and Table II, an O 1s – Li 1s ΔBE value is reported for Li metal, because even after extensive sputter-cleaning, a small amount of Li$_2$O is always detected. The results in Fig. 7 demonstrate that, while absolute BE values vary widely, the ΔBE values presented in this work are consistent with many literature reports. The dashed lines in Fig. 7 represent ±0.3 eV from the ΔBE values measured in this study (~9σ based on the error analysis from our measurements, shown in Table II).

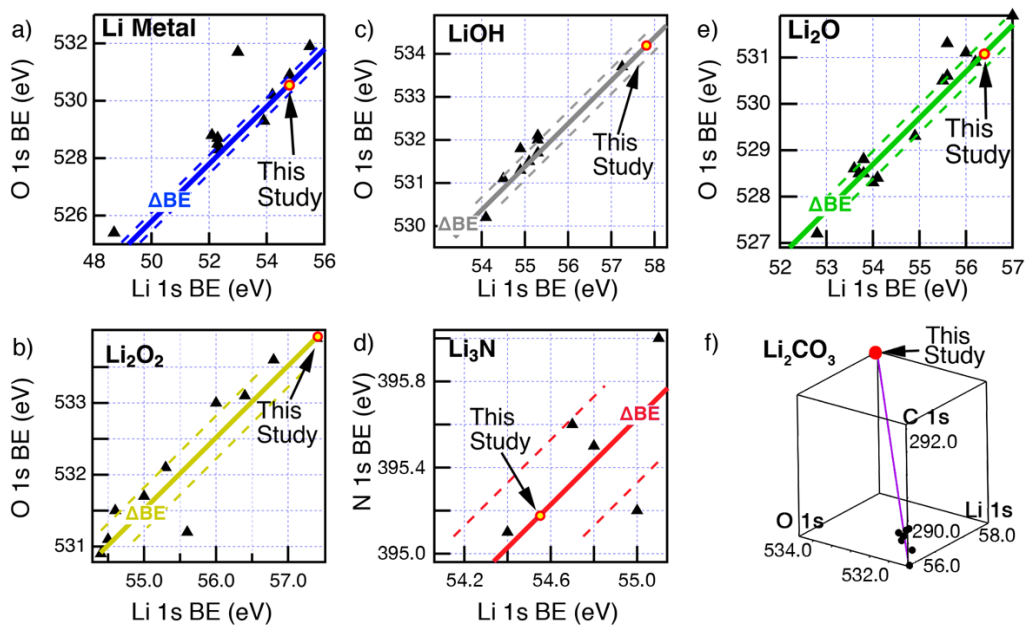

**Fig. 7:** *Plots of ΔBE values determined in this studied compared to other phases assignments in literature for a) Li$^0$ b) Li$_2$O$_2$ c) LiOH d) Li$_3$N e) Li$_2$O and f) Li$_2$CO$_3$.*

It is worth emphasizing that, because the ΔBE values are so similar for Li$_2$O$_2$ and LiOH, and the Li/O ratio is identical, the analysis summarized in Fig. 7 is generally not sufficient to definitively differentiate between these phases. Therefore, it is recommended that whenever possible valence-band spectra are also used to help identify these phases. Even so, it should be noted that co-existence of multiple phases makes interpretation of valence-band spectra challenging. Ideally, valence-band spectra can be acquired from well-characterized reference samples of phases believed to be present, and these can be used as the basis for curve-fitting unknown valence-band spectra.

It should also be noted that the methodology described thus far has relied on the implicit assumption that phases are distributed homogeneously within the detection volume, although clearly this is not necessarily always the case. In situations where phases are distributed non-uniformly, particularly in a layered structure, it might be necessary to modify sensitivity factors. To take a concrete example, if a thin layer of LiF were covered uniformly by a 3-nm-thick layer of other Li-containing phases, then we estimate that differences in electron inelastic mean free path (IMFP) values (nominally $\lambda_m^{Li\ 1s} \sim 2$ nm and $\lambda_m^{F\ 1s} \sim 1.5$ nm) would attenuate the F 1s signal significantly more than the Li 1s signal from this layer, by a factor of ~1.6.[54] In a curve-fitting analysis, this could lead to overestimation of Li content in the LiF layer and underestimation of Li content in the overlayer. A consequence of this could either be incorrect phase assignments, indeterminate curve-fitting results, or both. Therefore, it is imperative to be aware of these potential effects, to use complementary information about sample structure and morphology whenever possible, and to adjust sensitivity factors appropriately as required.

A related consideration is that the Li 1s sensitivity factor is the lowest of any element detectable by XPS. Therefore, in order to acquire Li 1s spectra that can support quantitative peak-fitting analysis, care must be taken to integrate long enough achieve acceptable signal-to-noise ratios.

To demonstrate an application of the aforementioned XPS analysis procedures in a real battery system, *operando* XPS measurements were performed on a $Li_2S-P_2S_5$ (LPS) solid-electrolyte symmetric cell (Li / LPS / Li), using the approach described in a recent study.[4] In this experiment, XPS measurements were performed on a sample initially comprised of an LPS pellet pressed onto Li foil. Application of an *in situ* current bias (constant current using the conditions described in Ref. 51) resulted first in the formation of an SEI at the exposed LPS surface, followed by plating of metallic lithium on the free surface. The current bias was then reversed, causing the plated Li to be driven back toward the opposing electrode. The resulting SEI / LPS / Li$_{foil}$ device structure was subjected to one additional electrochemical charge-discharge cycle. Subsequently, during the third charge cycle the Li / SEI interface was probed periodically by XPS. The resulting data set (Fig. 8a) demonstrates that Li$_2$O, Li$_2$S, and Li$_2$O$_2$ all exist in the exposed SEI at the start of the third charge cycle. As the charge cycle proceeds (t = 4 hr and t = 8 hr), plating of Li$^0$ is observed on the SEI surface, and the XPS spectra from the SEI phases evolve in two distinct ways. First, SEI peaks are progressively attenuated as Li$^0$ plates above them. And second, the overall resistivity of the cell appears to increase, as evidenced by shifts in absolute BEs of all SEI phases. However, as can be seen in Fig. 8 and Table III, the elemental ratios and ΔBE values for each phase show minimal changes.

This example serves to demonstrate that ΔBE and elemental ratios from XPS data can be used in a real system (a solid electrolyte in this example) to correctly identify phases (Table III), even when BE values shift over the course of an experiment. Moreover, this analysis also demonstrates that absolute BE shifts can provide valuable information about the sample being analyzed. In the present case, the observed BE shifts are due to the induced cell polarization in response to the *operando* current bias. We note that, as expected, the observed cell polarization disappeared when the *operando* current bias was removed. The observed increase in BEs indicate that cell polarization, and hence net cell resistivity, increases as galvanostatic charging proceeds. There are three broad possibilities for explaining the origin of the increased resistance: 1) an increase in the resistivity of the buried LPS / Li$_{foil}$ interface; 2) an increase in the resistivity of the LPS pellet; or 3) an increase in resistivity of the exposed SEI. Further measurements, beyond the scope of the present study, are needed to distinguish between these possibilities. Nevertheless, a detailed analysis of the formation and evolution of the SEI in a similar sample during the first charge–discharge cycle can be found in Ref. 51.

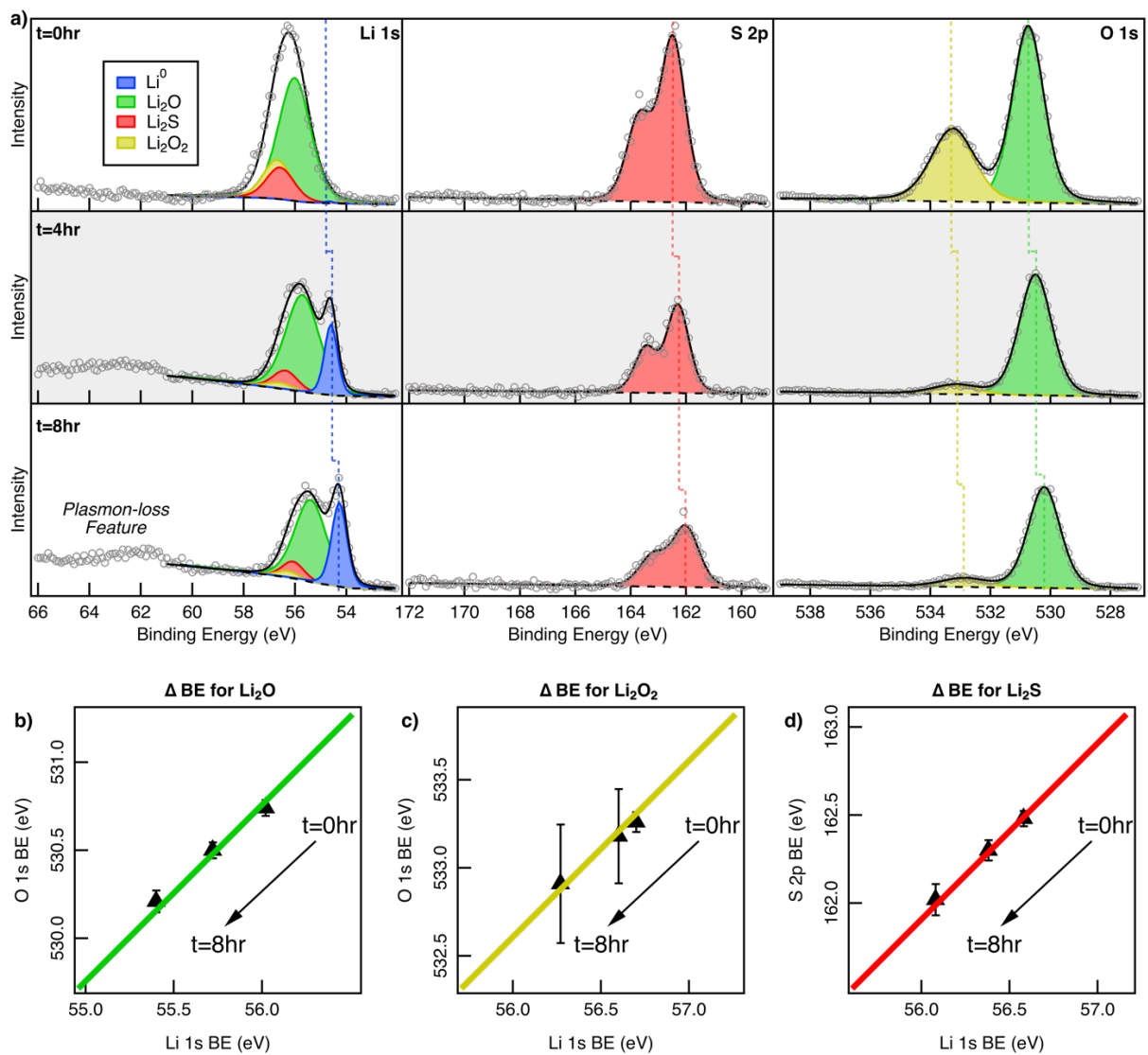

**Fig. 8: a)** *a) Spectral decomposition of Li 1s, O 1s, and S 2p for a Li/Li$_2$S-P$_2$S$_5$/Li solid electrolyte sample at three different points (t=0, 4, 8 hrs) during the third charge-discharge cycle. The graphs in b)-d) show that ΔBE values are constant for each phase even though the absolute BEs shift by different amounts during charging.*

**Table III:** *Absolute BE values for the spectral decomposition shown in Fig. 8. The ΔBE and elemental ratio for each phase are shown in the lower portion of the table*

|     |        | Binding Energy (eV) | | | | | |
| --- | ------ | --------- | --------- | --------- | --------- | --------- | --------- |
|     |        | t = 0 hr  |           | t = 4 hr  |           | Tt = 8 hr |           |
| Li  | Li$^0$ | 54.68     |           | 54.60     |           | 54.28     |           |
|     | Li$_2$O | 56.05    |           | 55.72     |           | 55.40     |           |
|     | Li$_2$S | 56.58    |           | 56.38     |           | 56.08     |           |
|     | Li$_2$O$_2$ | 56.70 |          | 56.60     |           | 56.27     |           |
| O   | Li$_2$O | 530.74   |           | 530.50    |           | 530.21    |           |
|     | Li$_2$O$_2$ | 533.26 |         | 533.18    |           | 532.91    |           |
| S   | Li$_2$S | 162.48   |           | 162.30    |           | 162.02    |           |
|     |        | ΔBE (eV)  | Ratio (Li:X) | ΔBE (eV) | Ratio (Li:X) | ΔBE | Ratio (Li:X) |
|     | Li$_2$O | 474.7    | 1.8       | 474.8     | 2.2       | 474.8     | 2.2       |
|     | Li$_2$O$_2$ | 476.6 | 0.9      | 476.6     | 1.1       | 476.6     | 1.1       |
|     | Li$_2$S | 105.9    | 1.5       | 105.9     | 1.7       | 105.9     | 1.8       |

CONCLUSIONS

In this work XPS was used to study eight different Li-battery relevant compounds. Analysis of data available in the literature revealed that a major source variability in BE assignments for electrically insulating phases commonly found in the battery literature is associated with charging due to the XPS measurement. We demonstrate that characteristic ΔBE values for each phase are unaffected by sample charging, and are much more reliable than absolute BEs for identifying phases. Therefore, we propose a methodology for XPS data analysis where characteristic ΔBE values are determined for relevant phases. These ΔBE values, along with appropriate elemental ratios, provide constraints during subsequent curve fitting of XPS spectra from battery samples that inherently account for sample charging and other effects that shift BEs on electronically insulating battery relevant materials. In this specific study, Li 1s and O 1s core levels we tracked during in-situ oxygen dosing of Li$^0$ to form Li$_2$O, thereby providing an absolute (uncharged) core level positions for Li$_2$O. This enabled the subsequent use of Li$_2$O core levels as an internal reference for the other compounds examined in this study, enabling accurate absolute core level positions. This method also effectively compensates for sample-to-sample and lab-to-lab variations caused by charging. Combining ΔBE values with elemental ratios and valene-band spectra provides a means for accurately identifying phases using XPS analysis, even spectra from difficult to analyze battery samples, where charging is a significant issue and multiple overlapping peak are common. The focus of the present study has been to apply these principles to common Li-containing inorganic phases, many of which are believed to be components of SEIs. Further work will need to be completed to extend these concepts to organic-containing SEI phases. And it important to note that as the complexity of a sample increases, there certainly will be cases where XPS measurements alone cannot unambiguously characterize the near-surface phase compositions. In such cases, information from complementary structural, compositional, and chemical characterization techniques will be indispensable.

AUTHOR INFORMATION

**Corresponding Author**

*Glenn Teeter: glenn.teeter@nrel.gov

**Funding Sources**

This work was supported by the U.S. Department of Energy under Contract No. DE-AC36-08GO28308 with the National Renewable Energy Laboratory. Funding was provided by the U.S. DOE Office of Energy Efficiency and Renewable Energy, Vehicle Technologies Office.

ACKNOWLEDGMENT

This work was supported by the U.S. Department of Energy under Contract No. DE-AC36-08GO28308 with the National Renewable Energy Laboratory. Funding was provided by the U.S. DOE Office of Energy Efficiency and Renewable Energy, Vehicle Technologies Office.

ORCID

KNW: 0000-0002-4049-7766

ABBREVIATIONS

solid-electrolyte interphase (SEI); X-ray photoelectron spectroscopy (XPS); lithium metal ($Li^0$); lithium oxide ($Li_2O$); lithium peroxide ($Li_2O_2$); lithium hydroxide (LiOH); lithium carbonate ($Li_2CO_3$); lithium nitride ($Li_3N$); binding energy (BE); Li-ion battery (LIB); nuclear magnetic resonance (NMR); ultra-high vacuum (UHV); Langmuirs (L); BE difference (ΔBE); valence-band maximum (VBM).

REFERENCES

(1)   Aurbach, D. A Short Review of Failure Mechanisms of Lithium Metal and Lithiated Graphite Anodes in Liquid Electrolyte Solutions. *Solid State Ionics* **2002**, *148* (3–4), 405–416.

(2)   Zhang, X.; Kostecki, R.; Richardson, T. J.; Pugh, J. K.; Ross, P. N. Electrochemical and Infrared Studies of the Reduction of Organic Carbonates. *J. Electrochem. Soc.* **2001**, *148* (12), A1341.

(3)   Zhuang, G. V.; Chen, Y.; Ross, P. N. The Reaction of Lithium with Dimethyl Carbonate and Diethyl Carbonate in Ultrahigh Vacuum Studied by X-Ray Photoemission Spectroscopy. *Langmuir* **1999**, *15* (3), 1470–1479.

(4)   Wood, K. N.; Steirer, K. X.; Hafner, S. E.; Ban, C.; Santhanagopalan, S.; Lee, S.-H.; Teeter,

G. Operando X-Ray Photoelectron Spectroscopy of Solid Electrolyte Interphase Formation and Evolution in Li2S-P2S5 Solid-State Electrolytes. *Nat. Commun.* **2018**, *9* (1), 2490.

(5) Cheng, X.-B.; Zhang, R.; Zhao, C.-Z.; Wei, F.; Zhang, J.-G.; Zhang, Q. A Review of Solid Electrolyte Interphases on Lithium Metal Anode. *Adv. Sci.* **2016**, *3* (3), 1500213.

(6) Wood, K. N.; Noked, M.; Dasgupta, N. P. Lithium Metal Anodes: Toward an Improved Understanding of Coupled Morphological, Electrochemical, and Mechanical Behavior. *ACS Energy Lett.* **2017**, *2* (3), 664–672.

(7) Kominato, A.; Yasukawa, E.; Sato, N.; Ijuuin, T.; Asahina, H.; Mori, S. Analysis of Surface Films on Lithium in Various Organic Electrolytes. *J. Power Sources* **1997**, *68* (2), 471–475.

(8) Xu, W.; Wang, J.; Ding, F.; Chen, X.; Nasybulin, E.; Zhang, Y.; Zhang, J.-G. Lithium Metal Anodes for Rechargeable Batteries. *Energy Environ. Sci.* **2014**, *7* (2), 513–537.

(9) Lin, C.-F.; Kozen, A. C.; Noked, M.; Liu, C.; Rubloff, G. W. ALD Protection of Li-Metal Anode Surfaces - Quantifying and Preventing Chemical and Electrochemical Corrosion in Organic Solvent. *Adv. Mater. Interfaces* **2016**, 1600426.

(10) Wood, K. N.; Kazyak, E.; Chadwick, A. F.; Chen, K.-H.; Zhang, J.-G.; Thornton, K.; Dasgupta, N. P. Dendrites and Pits: Untangling the Complex Behavior of Lithium Metal Anodes through Operando Video Microscopy. *ACS Cent. Sci.* **2016**, *2* (11), 790–801.

(11) Chen, K.-H.; Wood, K. N.; Kazyak, E.; LePage, W. S.; Davis, A. L.; Sanchez, A. J.; Dasgupta, N. P. Dead Lithium: Mass Transport Effects on Voltage, Capacity, and Failure of Lithium Metal Anodes. *J. Mater. Chem. A* **2017**, *5* (23), 11671–11681.

(12) Schroeder, M. a; Kumar, N.; Pearse, A. J.; Liu, C.; Lee, S. B.; Rubloff, G. W.; Leung, K.; Noked, M. DMSO–Li$_2$O$_2$ Interface in the Rechargeable Li–O$_2$ Battery Cathode: Theoretical and Experimental Perspectives on Stability. *ACS Appl. Mater. Interfaces* **2015**, *7* (21), 11402–11411.

(13) Yao, K. P. C.; Kwabi, D. G.; Quinlan, R. a.; Mansour, a. N.; Grimaud, A.; Lee, Y.-L.; Lu, Y.-C.; Shao-Horn, Y. Thermal Stability of Li2O2 and Li2O for Li-Air Batteries: In Situ

XRD and XPS Studies. *J. Electrochem. Soc.* **2013**, *160* (6), A824–A831.

(14) Landau, M. .; Gutman, A.; Herskowitz, M.; Shuker, R.; Bitton, Y.; Mogilyansky, D. The Role and Stability of Li2O2 Phase in Supported LiCl Catalyst in Oxidative Dehydrogenation of N-Butane. *J. Mol. Catal. A Chem.* **2001**, *176* (1–2), 127–139.

(15) Tonti, D.; Pettenkofer, C.; Jaegermann, W. Origin of the Electrochemical Potential in Intercalation Electrodes: Experimental Estimation of the Electronic and Ionic Contributions for Na Intercalated into TiS2. *J. Phys. Chem. B* **2004**, *108* (41), 16093–16099.

(16) Maibach, J.; Lindgren, F.; Eriksson, H.; Edström, K.; Hahlin, M. Electric Potential Gradient at the Buried Interface between Lithium-Ion Battery Electrodes and the SEI Observed Using Photoelectron Spectroscopy. *J. Phys. Chem. Lett.* **2016**, *7* (10), 1775–1780.

(17) Lindgren, F.; Rehnlund, D.; Källquist, I.; Nyholm, L.; Edström, K.; Hahlin, M.; Maibach, J. Breaking Down a Complex System: Interpreting PES Peak Positions for Cycled Li-Ion Battery Electrodes. *J. Phys. Chem. C* **2017**, *121* (49), 27303–27312.

(18) Oswald, S. Binding Energy Referencing for XPS in Alkali Metal-Based Battery Materials Research (I): Basic Model Investigations. *Appl. Surf. Sci.* **2015**, *351*, 492–503.

(19) Oswald, S.; Hoffmann, M.; Zier, M. Peak Position Differences Observed during XPS Sputter Depth Profiling of the SEI on Lithiated and Delithiated Carbon-Based Anode Material for Li-Ion Batteries. *Appl. Surf. Sci.* **2017**, *401*, 408–413.

(20) POVEY, A. F.; SHERWOO, P. M. A. Covalent Character of Lithium Compounds Studied by X-Ray Photoelectron Spectroscopy. *J. Chem. Soc., Faraday Trans.* **1974**, *2* (70), 1240–1246.

(21) Contour, J. P.; Salesse, A.; Froment, M.; Garreau, M.; Thevenin, J.; D. Warin. No Title. *J. Microsc. Spectrosc. électroniques* **1979**, *4*, 483.

(22) Surface State Analysis without Exposure to Atmosphere "XPS, AES" http://www.jfe-tec.co.jp/en/battery/case/09.html (accessed Feb 1, 2018).

(23) An, S. J.; Li, J.; Sheng, Y.; Daniel, C.; Wood, D. L. Long-Term Lithium-Ion Battery Performance Improvement via Ultraviolet Light Treatment of the Graphite Anode. *J. Electrochem. Soc.* **2016**, *163* (14), A2866–A2875.

(24) Ismail, I.; Noda, A.; Nishimoto, A.; Watanabe, M. XPS Study of Lithium Surface after Contact with Lithium-Salt Doped Polymer Electrolytes. *Electrochim. Acta* **2001**, *46* (10–11), 1595–1603.

(25) Cheng, L.; Crumlin, E. J.; Chen, W.; Qiao, R.; Hou, H.; Franz Lux, S.; Zorba, V.; Russo, R.; Kostecki, R.; Liu, Z.; et al. The Origin of High Electrolyte–electrode Interfacial Resistances in Lithium Cells Containing Garnet Type Solid Electrolytes. *Phys. Chem. Chem. Phys.* **2014**, *16* (34), 18294–18300.

(26) Kanamura, K.; Tamura, H.; Takehara, Z. XPS Analysis of a Lithium Surface Immersed in Propylene Carbonate Solution Containing Various Salts. *J. Electroanal. Chem.* **1992**, *333* (1–2), 127–142.

(27) Dedryvère, R.; Gireaud, L.; Grugeon, S.; Laruelle, S.; Tarascon, J. M.; Gonbeau, D. Characterization of Lithium Alkyl Carbonates by X-Ray Photoelectron Spectroscopy: Experimental and Theoretical Study. *J. Phys. Chem. B* **2005**, *109* (33), 15868–15875.

(28) Kanamura, K.; Tamura, H.; Shiraishi, S.; Takehara, Z.-I. XPS Analysis for the Lithium Surface Immersed in γ-Butyrolactone Containing Various Salts. *Electrochim. Acta* **1995**, *40* (7), 913–921.

(29) Shchukarev, A.; Korolkov, D. XPS Study of Group IA Carbonates. *Open Chem.* **2004**, *2* (2).

(30) Suo, L.; Hu, Y.-S.; Li, H.; Armand, M.; Chen, L. A New Class of Solvent-in-Salt Electrolyte for High-Energy Rechargeable Metallic Lithium Batteries. *Nat. Commun.* **2013**, *4*, 1481.

(31) Ota, H.; Sakata, Y.; Wang, X.; Sasahara, J.; Yasukawa, E. Characterization of Lithium Electrode in Lithium Imides/Ethylene Carbonate and Cyclic Ether Electrolytes. *J. Electrochem. Soc.* **2004**, *151* (3), A437.

(32) Lu, Y.-C.; Crumlin, E. J.; Veith, G. M.; Harding, J. R.; Mutoro, E.; Baggetto, L.; Dudney, N. J.; Liu, Z.; Shao-Horn, Y. In Situ Ambient Pressure X-Ray Photoelectron Spectroscopy Studies of Lithium-Oxygen Redox Reactions. *Sci. Rep.* **2012**, *2* (715), 1–6.

(33) Yen, S. P. S. Chemical and Morphological Characteristics of Lithium Electrode Surfaces. *J. Electrochem. Soc.* **1981**, *128* (7), 1434.

(34) Mallinson, C. F.; Castle, J. E.; Watts, J. F. Analysis of the Li KLL Auger Transition on Freshly Exposed Lithium and Lithium Surface Oxide by AES. *Surf. Sci. Spectra* **2013**, *20* (1), 113–127.

(35) Aurbach, D.; Pollak, E.; Elazari, R.; Salitra, G.; Kelley, C. S.; Affinito, J. On the Surface Chemical Aspects of Very High Energy Density, Rechargeable Li–Sulfur Batteries. *J. Electrochem. Soc.* **2009**, *156* (8), A694.

(36) Wu, Q.; Thissen, A.; Jaegermann, W. Photoelectron Spectroscopic Study of Li Oxides on Li over-Deposited V2O5 Thin Film Surfaces. *Appl. Surf. Sci.* **2005**, *250*, 57–62.

(37) Marino, C.; Darwiche, A.; Dupré, N.; Wilhelm, H. A.; Lestriez, B.; Martinez, H.; Dedryvère, R.; Zhang, W.; Ghamouss, F.; Lemordant, D.; et al. Study of the Electrode/Electrolyte Interface on Cycling of a Conversion Type Electrode Material in Li Batteries. *J. Phys. Chem. C* **2013**, *117* (38), 19302–19313.

(38) Wagner, C. D.; Muilenberg, G. E. *Handbook of X-Ray Photoelectron Spectroscopy: A Reference Book of Standard Data for Use in X-Ray Photoelectron Spectroscopy*; Perkin-Elmer, 1979.

(39) Oswald, S.; Mikhailova, D.; Scheiba, F.; Reichel, P.; Fiedler, A.; Ehrenberg, H. XPS Investigations of Electrolyte/electrode Interactions for Various Li-Ion Battery Materials. *Anal. Bioanal. Chem.* **2011**, *400* (3), 691–696.

(40) N.Futamura; T.Ichikawa; Imanishi, N.; Takeda, Y.; Yamamoto, O. Lithium Nitride Formation on Lithium Metal. In *Honolulu PRiME*; The Electrochemical Society, 2012; p 1137.

(41) Sun, Y.; Li, Y.; Sun, J.; Li, Y.; Pei, A.; Cui, Y. Stabilized Li3N for Efficient Battery Cathode Prelithiation. *Energy Storage Mater*. **2017**, *6* (August 2016), 119–124.

(42) Park, K.; Yu, B. C.; Goodenough, J. B. Li3N as a Cathode Additive for High-Energy-Density Lithium-Ion Batteries. *Adv. Energy Mater*. **2016**, *6* (10), 1–7.

(43) Younesi, R.; Norby, P.; Vegge, T. A New Look at the Stability of Dimethyl Sulfoxide and Acetonitrile in Li-O2 Batteries. *ECS Electrochem. Lett*. **2014**, *3* (3), A15–A18.

(44) Ishiyama, S.; Baba, Y.; Fujii, R.; Nakamura, M.; Imahori, Y. Surficial Chemical States of Li,N Synthesized on Lithium Target for Boron Neutron Capture Therapy. *Mater. Trans*. **2014**, *55* (3), 539–542.

(45) Guéguen, A.; Novák, P.; Berg, J. XPS Study of the Interface Evolution of Carbonaceous Electrodes for Li-O 2 Batteries during the 1st Cycle. *J. Electrochem. Soc*. **2016**, *163* (13), A2545–A2550.

(46) Marchini, F.; Herrera, S.; Torres, W.; Tesio, A. Y.; Williams, F. J.; Calvo, E. J. Surface Study of Lithium–Air Battery Oxygen Cathodes in Different Solvent–Electrolyte Pairs. *Langmuir* **2015**, *31* (33), 9236–9245.

(47) Zavadil, K. R.; Armstrong, N. R. Surface Chemistries of Lithium: Detailed Characterization of the Reactions of H2S, SO2, and SO2Cl2 Using XPS and EELS. *Surf. Sci*. **1990**, *230* (1–3), 61–73.

(48) Wang, K.; Ross Jr., P. N.; Kong, F.; McLarnon, F. The Reaction of Clean Li Surfaces with Small Molecules in Ultrahigh Vacuum. I. Dixoygen. *J. Electrochem. Soc*. **1996**, *143* (2), 422–428.

(49) Khosravi, J. PRODUCTION OF LITHIUM PEROXIDE AND LITHIUM OXIDE IN AN ALCOHOL MEDIUM, McGill University, 2007.

(50) Seah, M. P.; Gilmore, I. S.; Beamson, G. XPS: Binding Energy Calibration of Electron Spectrometers 5--Re-Evaluation of the Reference Energies. *Surf. Interface Anal*. **1998**, *26* (9), 642–649.

(51) Schmid, M.; Steinrück, H. P.; Gottfried, J. M. A New Asymmetric Pseudo-Voigt Function for More Efficient Fitting of XPS Lines. *Surf. Interface Anal*. **2014**, *46* (8), 505–511.

(52) Brako, R.; Newns, D. M.; Lloyd, P.; Weightman, P. Many-Electron Singularity in X-Ray Photoemission and X-Ray Line Spectra from Metals Many-Electron Singularity I. **1970**.

(53) Kowalczyk, S. P.; Ley, L.; McFeely, F. R.; Pollak, R. A.; Shirley, D. A. X-Ray Photoemission from Sodium and Lithium. *Phys. Rev. B* **1973**, *8* (8), 3583–3585.

(54) Seah, M. P.; Dench, W. A. Quantitative Electron Spectroscopy of Surfaces : *Surf. Interface Anal*. **1979**, *1* (1), 2–11.